\documentclass[12pt]{article}
\usepackage{amsmath}
\usepackage{amssymb}
\usepackage{mathptmx}
\usepackage{amsfonts}
\usepackage[mathscr]{eucal}
\usepackage[dvips]{graphicx}
\usepackage{color}

\newcommand{\I}{\textrm{I}}

\newcommand{\bos}{\textrm{b}}

\newcommand{\ms}[1]{\ensuremath{\mathscr{#1}}}

\newcommand{\Fb}{\mathscr{F}_{\, \textrm{b}}  } 
\newcommand{\Fbfin}{\mathscr{F}_{\, \textrm{b} , \textrm{fin}}  }

\newcommand{\Dzero}{ \mathscr{D}_{0} }

\newcommand{\Rd}{\mathbb{R}^{\, d} }

\newcommand{\dx}{d \mathbf{x} }

\newcommand{\dk}{d \mathbf{k} }

\newcommand{\HI}{H_{\textrm{I}}}
\newcommand{\HIk }{H_{\textrm{I} } ( \kappa ) }
\newcommand{\Hk}{H (\kappa)  }

\newcommand{\SI}{S_{\textrm{I}}}

\newcommand{\Sk}{S ( \kappa ) }

\newcommand{\Nb}{N_{\textrm{b}}}

\newcommand{\chiI}{\chi_{\, \textrm{I}}  }
\newcommand{\chiIx}{\chi_{\, \textrm{I}} (\mathbf{x}) }

\newcommand{\chib}{\chi_{\, \textrm{b}}  }
\newcommand{\chibk}{\chi_{\, \textrm{b}} (\mathbf{k}) }

\newcommand{\omegak}{\omega  (\mathbf{k}) }

\newcommand{\Ezerok}{E_{0} ( \kappa  ) }

\newcommand{\Phik}{\Phi_\kappa   }
\newcommand{\PsiEk}{\Psi_{E_{\kappa}}   }
\newcommand{\Ek}{{E_\kappa} }

\newcommand{\mbf}[1]{\ensuremath{\mathbf{#1}}}
\newcommand{\mbb}[1]{\ensuremath{\mathbb{#1}}}

\newcommand{\sqzb}{\ensuremath{d\Gamma_{\textrm{b}}}}
\newcommand{\tens}{\otimes}
\newcommand{\ntens}{\otimes^{n}}
\newcommand{\nstens}{\otimes^{n}_{\textrm{s}}}

\newcommand{\restr}{\upharpoonright}

 \newcommand{\Omegazero}{\Omega_{0}}
 \newcommand{\Omegak}{\Omega_{\kappa }}

\newcommand{\Czeroinfty}{C_{0}^{\, \infty}}

\newcommand{\1}{{\small \text{1}}\hspace{-0.35em}1}

\newcommand{\rhob}{\rho_{ \textrm{b}}}
\newcommand{\rhobx}{\rho_{ \textrm{b} , \mathbf{x}}}

\newcommand{\phiS}{\phi_{\, \textrm{S}}}

\newtheorem{theorem}{Theorem}[section]
\newtheorem{proposition}[theorem]{Proposition}
\newtheorem{lemma}[theorem]{Lemma}
\newtheorem{corollary}[theorem]{Corollary}

\begin{document}
\begin{center}
{\LARGE The first order expansion of a ground state energy  of the $\phi^4$ model with cutoffs}
 \\
 $\;$ \\
 {\large Toshimitsu Takaesu }  \\
 $\;$ \\
\textit{Faculty of Science and Technology, Gunma University,\\ Gunma, 371-8510, Japan }
\end{center}

\begin{quote}
\textbf{Abstract} 
In this paper, we investigate the $\phi^4$ model with cutoffs.  By introducing  a spatial cutoff and   a momentum cutoff, the total Hamiltonian is a self-adjoint operator on a boson Fock space.  Under  regularity conditions of the momentum cutoff,  we obtain the first order expansion  of  a non-degenerate ground state energy of the total Hamiltonian. \\
\end{quote}
{\small
key words : Quantum field theory, Hilbert space, Self-adjoint operator}.

\section{Introduction}
In this paper we consider the $d$-dimensional $\phi^4$ model with cutoffs. The Hilbert space for the system is defined by a boson Fock space. The total Hamiltonian  is given by a sum of the free Hamiltonian
 and  perturbation
\begin{equation}
 \qquad \Hk \; = \; H_{0} + \kappa  \int_{\Rd} \chiIx \phi (\mbf{x})^4 \dx , \qquad \quad \kappa >0 .
\end{equation}
The  dispersion relation of    $H_0 $ is  $\omega (\mbf{k})= \sqrt{\mbf{k}^2 + m^2}$,  $m\geq 0$, and  a momentum cutoff is imposed on the field operator $\phi (\mbf{x})$. From the beginning of constructive quantum field theory, the $\phi^4$ model has been investigated  (\cite{GJ68}).   The main interest in this paper is  to investigate a  perturbative expansion of a  ground state energy of  $ \Hk $.
A mathematical feature of  the $\phi^4$ model is  the singular   perturbation, which means that the perturbation is  not relatively bounded to the free Hamiltonian (\cite{GJ69}).   
In particular, in the  case of $m=0$, the ground state energy of $H_0$ is an embedded eigenvalue. 
Hence   we cannot  apply the Kato perturbation theory (\cite{Kato,RS}).  
We  suppose that the  total  Hamiltonian has  a  non-degenerate ground state. 
In the main theorem, we   derive the first order expansion of the ground state energy. 
 
To prove the main theorem, we  apply  Arai's  new perturbation method \cite{Ar14},  which is based on the Brillouin-Wigner perturbation methods and applied to the generalized spin-boson model. 
In the proof, we derive an upper bound of the ground state energy and   a norm inequality, called  $H$-bound. We also  show  the pull-through formula and prove the boson number  bound,  which play an important role to prove  the existence and uniqueness of the ground states   for  the interaction systems of  massless Bose fields ( e.g., \cite{AH97, BFS98, BFS99, Ge00, GLL01, Hi05}).

The   asymptotic perturbations of the ground states  of massless quantum filed models have been investigated, and  refer to the Pauli-Fierz models  (\cite{FMS11, GH09,  HS02,HH11A, HH11B}), the spin-boson model  (\cite{BHL18, HH11C}) and references therein. For the existence of the ground states of singular perturbation models, refer to e.g. \cite{Ta10, Wa17}.  In addition, the asymptotic completeness  of the one-dimensional $\phi^4$ model with a spatial cutoff was shown  in \cite{DG00}.

This paper is organized as follows.
In Section 2, we give the definitions of the state space and total Hamiltonian, and state the main result.
In Section 3,  we derive  the upper bound of the ground state energy and the $H$-bound. By  the pull-through formula, we show the boson number bound. Applying  the general theory in \cite{Ar14}, we prove the main theorem.

\section{Main Result}
We define the state space and the total Hamiltonian  by means of Fock space theory  \cite{Arai}.
The Hilbert  space for the system is defined by
\[
\Fb  = \bigoplus_{n=0}^{\infty} \left( \nstens L^2 (\Rd ) \right) ,
\]
where $  \nstens \ms{X} $ denotes the n-fold symmetric tensor product of a Hilbert space $\ms{X}$ with $ \tens_{\textrm{s}}^0 \ms{X} = \mathbb{C} $.  The Fock vacuum is defined by $ \Omegazero = \{ 1 , 0,  0, \ldots \} \in \Fb $.
The creation operator is defined by
$ \left( a^{\dagger} (f) \Psi \right)^{(n)} = \sqrt{n} S_n (f \tens \Psi^{(n-1)})$, $n \geq 1 $,
and $ \left( a^{\dagger} (f) \Psi \right)^{(0)} =  0 $ where $S_n$ is the symmetrization operator on $ \ntens L^2 (\Rd ) $. The   annihilation  operator is defined by  $a(f) = (a^{\dagger} ( f) )^{\ast}   $  where   $X^{\ast}$  denotes the adjoint of $X$.   The finite particle space on a subspace $\ms{M} \subset  L^2 (\Rd )$ is defined by 
\[
 \Fbfin (\ms{M}) \,  = \,  \text{L.H.} \left\{  \Omegazero , a^{\dagger}(f_1) \ldots a^{\dagger}(f_n)\Omegazero  \left. \frac{}{} \right| f_1 , \ldots , f_{n}  \in \ms{M} , \, n \in \mbb{N} \right\}
\]
Creation  and annihilation operators satisfy the canonical commutation relations on a finite particle subspace $\Fbfin (\ms{M})$ ;
\begin{align}
& [a(f) , a^{\dagger} (g)] = (f,g)    , \label{CCR1}   \\
& [a(f) , a(g) ] =  [a^{\dagger}(f) , a^{\dagger}(g) ] =  0 .   \label{CCR2}
\end{align} 
The Segal field operator is defined by 
\[
\phiS (f) \; = \; \frac{1}{\sqrt{2}} \,  \overline{a(f) + a^{\dagger} (f)} ,
\] 
 where  $\overline{X}$  denotes the closure of   $X$. Let $\omega (\mbf{k}) = \sqrt{\mbf{k}^2 + m^2} $, $m \geq 0$.  
The free Hamiltonian is defined by 
 \[
 H_{0} \; = \; \sqzb (\omega ) ,
 \]
where $\sqzb (X)$ is the second quantization defined by 
$  \left(\sqzb (X)  \Psi \right)^{(n)} = \sum\limits_{j=1}  (\1 \tens \ldots  \tens \1 \tens
\underbrace{X}_{jth} \tens \1 \tens \cdots \tens  \1 ) \Psi^{(n)} $, $ n \geq 1 $ and $\left(\sqzb (X)  \Psi \right)^{(0)} =0$. The number operator is defined by $N_{\bos} = \sqzb (1) $. 
To define the interaction,  we introduce the ultraviolet cutoff  $ \chib $ and spatial cutoff $\chiI$.  Suppose the following conditions.
\begin{quote}
\textbf{(H.1;Ultra-violet cutoff)}   $ \| \chib \|_{L^2} < \infty ,
  \|  \frac{\chib }{\sqrt{\omega }^l } \|_{L^2}  < \infty , $ $l=1,2 $. \\
\textbf{(H.2;Spatial cutoff)}  $ \| \chiI \|_{L^1} < \infty   $. 
\end{quote}
Let 
\[
\phi (\mbf{x}) \; = \; \phiS ( \rhobx ) ,
\]
where $  \rhobx (\mbf{k})= \rhob (\mbf{k} ) e^{-i \mbf{k} \cdot \mbf{x}} $ with 
$\rhob (\mbf{k}) = \frac{\chibk }{\sqrt{\omega (\mbf{k)}}}$.  \\
The interaction is defined by 
\[
\HI \Psi  =   \int_{\Rd} \chiIx  \phi (\mbf{x})^4   \Psi  \dx  ,
\]
where the integral  is in the sense of  the strong Bochner integral and the domain is given by
$ \ms{D} (\HI ) = \left\{    \Psi  \in \Fb \left| \right.   \int_{\Rd} |  \chiIx | \, \|  \phi (\mbf{x})^4  \Psi  \|  \dx < \infty   \right\}$.

$\; $ \\
Let 
\[
\Hk \;    = \;   H_{0}  \, +\HIk  ,
\] 
where $\HIk =  \kappa \, \overline{  \HI  } $. 
It follows that  $ \Hk  $ is essentially self-adjoint on $\ms{D}_0 = \Fbfin (\Czeroinfty (\Rd ) )  $. 
We briefly give the proof in a similar way to   the singular perturbation models  \cite{Ta10, Wa17}. By (\cite{Ar91};Theorem 2.1),   $ \Hk $ is essentially self-adjoint on $\ms{D} (H_0) \cap  \ms{F}_{\bos , 0} $ where
\[
\ms{F}_{\bos , 0}  = \left\{ \Psi = \{ \Psi^{(n)}  \}_{n=0}^{\infty } \in \Fb
\left| \right. {}^{\exists} l \in \mbb{N} \;  \text{s.t.} \; {}^{\forall} l ' > l, \Psi^{(l')} = 0 \right\} .
\] 
Let $\Psi \in \ms{D} (H_0) \cap \ms{F}_{\bos , 0} $. Since $\ms{D}_0 $ is a core of $H_{0}$, there exists a sequence  $\{ \Psi_n \}_{n= 1}^{\infty}$ of $\ms{D}_0$ such that
$\lim\limits_{n\to \infty} \| \Psi_n - \Psi \| =0$ and $\lim\limits_{n\to \infty} \| H_0 \Psi_n - H_0 \Psi \| =0$. In addition $ H_{\textrm{I} } $ is bounded on $\nstens L^2 (\Rd )$, and hence  $\lim\limits_{n , m \to \infty} \| \HI \Psi_n - \HI \Psi_m \| =0$. Therefore   $\Psi \in \ms{D} (  \overline{  \HI  })  $ and we have $\Psi \in \ms{D} (H_0 ) \cap \ms{D}   (  \overline{  \HI  }) $.  

$\; $ \\
Let  $E_{0} = \inf \sigma (H_{0} )  $.   It is seen that and  $E_{0} =0 $ and  $H_0 \Omega_0 = E_{0}  \Omega_0 $.   
Let  $\Ezerok = \inf \sigma (\Hk )$.  Assume the condition below. 
\begin{quote}
\textbf{(H.3)} There exists $\kappa_{\ast} >0 $ such that for all $0< \kappa <\kappa_{\ast} $,   $   \Hk  $ has a ground state with dim ker $(\Hk - \Ezerok )=1$ . 
\end{quote}
 Let  $\Omegak $ be 
 the normalized ground state ;
\[
\qquad \quad 
\Hk \Omegak = \Ezerok \Omegak ,  \qquad \| \Omegak  \| = 1 .
\]
Suppose the additional condition below.  
\begin{quote}
\textbf{(H.4)}  $   \|\frac{\chib }{\omega^{3/2}} \|_{L^2} < \infty   $. 
\end{quote}

$\; $ \\
Here we state the main theorem .

\begin{theorem}  \label{MainTheorem}
Suppose \textbf{(H.1)} - \textbf{(H.4)}. Then 
\[
\Ezerok  = \kappa ( \Omegazero , \HI  \Omegazero )  + o (\kappa ).
\]
\end{theorem}




\section{Proof of Main Theorem}
\subsection{Upper Bound of  Ground State Energy}
Let $P_0$ be the projection onto the closed subspace $ \ms{M}_0 = \left\{ z  \Omega_0 \left| \frac{}{} \right. z \in \mbb{C} \right\}$.  Let $P_0^{\bot} = \1 - P_0 $.  
It follows that $H_{0}$ is reduced by $ \ms{M}_0 $ and $\ms{M}_0^{\bot}  $, respectively. Let
$ H_{0}^{\bot} = H_{0  \, \restr \ms{D}(H_0) \cap  \ms{M}_0^{\bot}}$ where $ X_{\restr \ms{M}}$ denotes the restriction of $X$ to a subspace $\ms{M} $.

\begin{lemma} \label{11/25l1} Suppose \textbf{(H.1)} - \textbf{(H.2)} and \textbf{(H.4)}.  Then for all $\kappa \in \mbb{R}$, 
\[
\Ezerok  \leq  \frac{1}{1 + \nu_0 }
\left(  ( \Omegazero , \HI \Omegazero ) \kappa - a \kappa^2 + b \kappa^3 \right) ,
\]
where 
\begin{align*}
& \nu_0 = \|  (H_0^{\bot}  )^{-1}  P_0^{\bot} \HI  \Omegazero \|^2 , \\ 
&  a = (P_0^{\bot} \HI \Omegazero ,  (H_0^{\bot}  )^{-1}  P_0^{\bot} \HI \Omegazero )  , \\
& b=( (H_0^{\bot}  )^{-1} P_0^{\bot} \HI \Omegazero , \HI (H_0^{\bot})^{-1}  P_0^{\bot} \HI \Omegazero)  .
\end{align*}
\end{lemma}
Before proving the lemma,  we  review  basic properties of the creation operator and the second quantization. The creation operator acts the $n$-particle state  such as
\begin{equation}
\left(  a^{\dagger} (f) \Psi \right)^{(n)} (\mbf{k}_1 , \ldots , \mbf{k}_n)
= \frac{1}{\sqrt{n}}  \sum_{j=1}^{n} f(\mbf{k}_j )  \Psi^{(n-1)} (\mbf{k}_1 , \ldots , \tilde{\mbf{k}}_{j} \ldots , \mbf{k}_n)  ,
\label{11/25b} 
\end{equation} 
where $\tilde{\mbf{k}}  $ stands for omitting the variable $\mbf{k}$. 
We also see that for all $\Psi \in \ms{D} ( (H_0^{\bot})^{-1} )$,
\begin{equation}
\qquad 
\left(  (H_0^{\bot})^{-1}  \Psi \right)^{(n)} (\mbf{k}_1 , \ldots , \mbf{k}_n)
= \frac{1}{ \sum\limits_{j=1}^{n} \omega ( \mbf{k}_{j})  }  \Psi^{(n)}(\mbf{k}_1 , \ldots , \mbf{k}_n)    , \qquad n \geq 1,
\label{11/25c} 
\end{equation}
and $ \left(  (H_0^{\bot})^{-1}  \Psi \right)^{(0)} = 0 $.

$\;$ \\
\textbf{(Proof of Lemma  \ref{11/25l1})}
By  Proposition A in Appendix, it is enough to show that 
\begin{equation}
\Omegazero \in \ms{D} (\HI (H_0^{\bot}  )^{-1} P_0^{\bot} \HI ) .  \label{11/25a} 
\end{equation}
Let 
$ \Phi = \phiS (f)^4 \Omega_0 $, $f \in \ms{D} (\omega^{-1})$. 
 By the canonical commutation relations (\ref{CCR1}), (\ref{CCR2}) and $a(f) \Omega_{0}= 0$, 
 we see that  $P_0^{\bot} \Phi$ is the form $P_0^{\bot} \Phi = \sum\limits_{l=1}^{4} \lambda_l (f)a^{\dagger } (f )^l  \Omegazero  $.  Note that $f \in \ms{D} (\omega^{-1})$. Then  by (\ref{11/25b}) , (\ref{11/25c}) and  $\frac{1}{ \sum_{j=1}^{n} (\omega ( \mbf{k}_{j}))} \ \leq  \frac{1}{  (\omega ( \mbf{k}_{l}))} $, $l=1 , \ldots , n$  it follows that  $ \Phi  \in \ms{D} ((H_{0}^{\bot} )^{-1})$.
In addition $(H_0^{\bot})^{-1} $ maps $n$-particle states to $n$-particle states, and hence 
$ (H_0^{\bot})^{-1} \Phi  \in \ms{F}_{\bos , 0} $.  This  concludes that    (\ref{11/25a}) follows.
$\square $

  
\subsection{H-bound}
Let  $f \in \ms{D} (\omega^{-1/2})$. 
 For all  $\Psi \in \ms{D} ( H_{0}^{1/2} )$, 
\begin{align}
& \| a (f) \Psi \| \leq  \| \frac{f}{ \sqrt{\omega}}  \|  \| H_0^{1/2} \Psi \| ,   \label{azerobound} \\
& \| a^{\dagger} (f) \Psi \| \leq    \| \frac{f}{ \sqrt{\omega}}  \|  \| H_0^{1/2} \Psi \| 
+  \|f \| \,  \| \Psi \|  .  \label{czerobound}
\end{align}
By (\ref{azerobound}) and (\ref{czerobound}),
\begin{equation}
\| \phiS (f) \Psi  \| \leq \sqrt{2}  \| \frac{f}{ \sqrt{\omega}}  \|  \| H_0^{1/2} \Psi \|
+\frac {\|f \|}{\sqrt{2}} \,  \| \Psi \|  .  \label{phiSdbound}
\end{equation}
Let  $f \in \ms{D} (\omega)$. Then for all $\Phi \in \ms{D}_0 $, 
\begin{align}
& [a(f) , H_0 ] \Phi  = a(\omega f) \Phi ,   \label{commazero} \\
& [a^{\dagger}(f) , H_0 ] \Phi  = -a^{\dagger}(\omega f)  \Phi . \label{commaczero} 
\end{align}
By  (\ref{commazero}) and (\ref{commaczero}),
\begin{equation}
[\phiS (f ) , H_0 ]   \Phi = i \phiS ( i \omega f )  \Phi .  \label{commphizero} 
\end{equation}
The next lemma  is easily proven  by  (\ref{phiSdbound})  and (\ref{commphizero}).

\begin{lemma} \label{9/15a}  Let $f \in \ms{D} (\omega)$. Then, it holds that for all $\Phi \in \ms{D}_0 $, 
\begin{align*}
 & \textbf{(i)} \;    [ \phiS (f)^2 , [\phiS (f)^2 , H_0 ]]\Phi = - 4 (f, \omega f ) \phiS (f)^2 \Phi  , \\ 
& \textbf{(ii)} \; 
\left| \left( \Phi , \, [  \phiS (f)^2 , [\phiS (f)^2 , H_0 ]]\Phi \right) \right|  \; \leq \;
  4 \| \omega^{1/2} f  \|^2  \left( 4 \| \frac{f}{ \sqrt{\omega}}  \|^2   \| H_0^{1/2} \Phi \|^2 
+ \|f \|^2  \,  \|  \Phi \|^2  \right) .
\end{align*}
\end{lemma}

\begin{proposition} \label{11/24p1}  
Suppose \textbf{(H.1)} and \textbf{(H.2)}. Let $\epsilon >0$. Then it holds that for all $\Psi \in \ms{D} (\Hk ) $, \\
\begin{equation}
(1-c_{\bos} \epsilon \kappa ) \|  H_0 \Psi  \|^2   + \|  \HIk \Psi  \|^2 \,  \leq \,   \| \Hk \Psi  \|^2   +  \left (4 d_{\bos} + \frac{c_{\bos}}{4 \epsilon }    \right)  \kappa     \|  \Psi  \|^2   .
  \label{11/24a}
\end{equation}
where $c_{\bos}  = 16 \| \chiI \|_{L^1}  \| \chib  \|^2 \| \frac{\chib}{ \omega}  \|^2 $ and 
$d_{\bos}  =  \| \chiI \|_{L^1}  \| \chib  \|^2 \| \frac{\chib}{ \sqrt{\omega}}  \|^2 $.

\end{proposition}
\textbf{(Proof)} Let $\Phi \in \Dzero $. It is seen that
\begin{equation}
\|  \Hk   \Phi  \|^2  \,  = \,  \|  H_0 \Phi  \|^2   
+ \kappa  (\Phi , (H_0 \HI + \HI H_0 ) \Phi ) 
+ \kappa^2  \|  \HI \Phi  \|^2  .   \label{11/24b}
\end{equation}
By $ X Y^2   + Y^2 X = 2YXY + [Y , [Y, X]] $, we have
\begin{align*}
& (\Phi , (H_0 \HI + \HI H_0 ) \Phi ) \\
& =  \int_{\Rd} \chiI (\mbf{x}) (\Phi , (H_0 \phi (\mbf{x})^4 + \phi  (\mbf{x})^4 H_0 ) \Phi )  \dx   \\
& = 2 \int_{\Rd} \chiI (\mbf{x}) (\Phi , ( \phi (\mbf{x})^2  H_0 \phi  (\mbf{x})^2  ) \Phi )  \dx   
  + \int_{\Rd} \chiI (\mbf{x}) (\Phi ,  [ \phi (\mbf{x})^2, [\phi (\mbf{x})^2, H_0  ] ]\Phi )  \dx   \\
&\geq \int_{\Rd} \chiI (\mbf{x}) (\Phi ,  [ \phi (\mbf{x})^2, [\phi (\mbf{x})^2, H_0  ] ]\Phi ) \dx .
\end{align*}
By Lemma \ref{9/15a} \textbf{(ii)}, it follows that 
\begin{align*}
& \left| \int_{\Rd} \chiI (\mbf{x}) (\Phi ,  [ \phi (\mbf{x})^2, [\phi (\mbf{x})^2, H_0  ] ]\Phi ) \dx \right| \\
&\qquad \qquad  \quad  \leq  4 \| \chiI  \|_{L^1}  \| \omega^{1/2} \rhob  \|^2  \left( 4 \| \frac{\rhob}{ \sqrt{\omega}}  \|^2   \| H_0^{1/2} \Phi \|^2 
+ \| \rhob \|^2  \,  \|  \Phi \|^2  \right) .
\end{align*}
Thus we have
\begin{equation}
 (\Phi , (H_0 \HI + \HI H_0 ) \Phi )
\geq - \left( c_{\bos}   \|  H_0^{1/2}  \Phi \|^2   +  4 d_{\bos}   \| \Phi \|^2  \right), \notag 
\end{equation}
where $c_{\bos}  = 16 \| \chiI \|_{L^1}  \| \chib  \|^2 \| \frac{\chib}{ \omega}  \|^2 $ and 
$d_{\bos}  =  \| \chiI \|_{L^1}  \| \chib  \|^2 \| \frac{\chib}{ \sqrt{\omega}}  \|^2 $.
By the inequality  $\|  H_0^{1/2}  \Phi \|^2 \leq \epsilon \|   H_0 \Phi \|^2 + \frac{1}{4 \epsilon } \| \Phi  \|^2  $, $\epsilon >0 $, we have
\begin{equation}
 (\Phi , (H_0 \HI + \HI H_0 ) \Phi )
\geq - c_{\bos}  \epsilon   \|  H_0^{1/2}  \Phi \|^2   - 
\left (4 d_{\bos} + \frac{c_{\bos}}{4 \epsilon }    \right)  \| \Phi \|^2   .
   \label{11/24c}
\end{equation}
By (\ref{11/24b}) and (\ref{11/24c}), we have
\begin{equation}
\|  \Hk   \Phi  \|^2  \,  \geq  \,  (1- c_{\bos} \epsilon \kappa  ) \|  H_0 \Phi  \|^2   
 -   \left (4 d_{\bos} + \frac{c_{\bos}}{4 \epsilon }    \right) \kappa   \| \Phi \|^2  + \kappa^2  \|  \HI \Phi  \|^2  .  \label{11/24d}
\end{equation}
Since $\ms{D}_0$ is the core of $\Hk$, we see that (\ref{11/24d}) holds for all $ \Psi \in \ms{D} (\Hk )$. Thus the proof is complete. $\square $

\begin{corollary} \label{11/24c1}  (H-bound)
Suppose \textbf{(H.1)} and \textbf{(H.2)}. Then, it holds that for all $\Psi \in \ms{D} (\Hk ) $ and for all $\epsilon >0$ such that $ \epsilon < \frac{1}{c_{\bos} \kappa }   $, \\
\begin{equation}
 \|  H_0 \Psi  \|^2   +   \|  \HIk \Psi  \|^2 \,  \leq \,  \lambda_{\epsilon , \kappa} \| \Hk \Psi  \|^2   +   \mu_{\epsilon , \kappa}  \|  \Psi  \|^2   .
\end{equation}
where $\lambda_{\epsilon, \kappa } = \frac{1}{1-c_{\bos}\epsilon \kappa}  
 $ and $\mu_{\epsilon , \kappa }= \frac{\kappa }{1-c_{\bos} \epsilon \kappa }
  \left (4 d_{\bos} + \frac{c_{\bos}}{4 \epsilon }    \right) $.
\end{corollary}

\subsection{Boson Number Bound}
We introduce the operator kernel of  the annihilation operator which satisfies that 
\[
(\Phi , a (f) \Psi )= \int_{\Rd } f(\mbf{k})^{\ast} (\Phi, a ( \mbf{k}) \Psi ) \dk,  \qquad \Phi , \Psi  \in \ms{D}(H_0 ) ,
\]
where $ (a ( \mbf{k}) \Psi)^{(n)} ( \mbf{k}_1 , \ldots , \mbf{k}_n ) 
= \sqrt{n+1 }  \Psi^{(n+1 )} (\mbf{k} ,  \mbf{k}_1 , \ldots , \mbf{k}_n ) $, $n = 0, 1, \ldots $.  \\
The  weak commutator of operator $X$ and $Y$ is defined by 
\[
[X ,Y ]^{0} (\Phi , \Psi ) = (X^{\ast}\Phi, Y \Psi ) -(Y^{\ast}\Phi, X \Psi ) ,
\]
for all  $\Phi \in \ms{D}(X^{\ast}) \cap \ms{D}(Y^{\ast})  $ and $ \Psi \in \ms{D}(X) \cap  \ms{D}(Y)$.

\begin{proposition} \label{11/24p2}  (Pull-Through Formula) 
Suppose \textbf{(H.1)} - \textbf{(H.3)}. 
Then
\[
a (\mbf{k}) \Omegak = - 2 \sqrt{2} \kappa \frac{\chibk }{{\sqrt{\omega (\mbf{k})}}}
(\Hk  -\Ezerok  + \omega (\mbf{k}) )^{-1}
\int_{\Rd} \chiIx e^{-i \mbf{k} \cdot \mbf{x}} \phi(\mbf{x})^3  \Omegak \dx , \; \;  \text{a.e.} \, \Rd ,
\]
where the integral in  the right-hand side is  the strong Bochner integral.
\end{proposition}
\textbf{(Proof)} Let $\Phi \in \ms{D}_0 $.  It is seen that
\begin{align}
[\Hk , a(f) ]^0 (\Phi ,  \Omegak  ) & =    (\Hk \Phi , a(f) \Omegak   ) -   ( a^{\dagger} (f) \Phi , \Hk  \Omegak   )  \notag \\
 &= ((\Hk - \Ezerok )\Phi , a(f)   \Omegak  )  .\label{11/24d1}
\end{align}
We also see that
\begin{align}
[\Hk , a(f) ]^0 (\Phi ,  \Omegak  )  &=   [H_0  , a(f) ]^0 (\Phi ,   \Omegak ) + \kappa [\HI , a(f) ]^0 (\Phi , \Omegak  )  \notag \\
&=  - (\Phi , a(\omega f )  \Omegak    ) + \kappa [\HI , a(f) ]^0 (\Phi ,   \Omegak ) . \label{11/24d2} 
\end{align}
By  (\ref{11/24d1}) and (\ref{11/24d2}), we have
\begin{equation}
 ((\Hk - \Ezerok )\Phi , a(f)   \Omegak )  +  (\Phi , a(\omega f )    \Omegak  ) = 
 \kappa [\HI , a(f) ]^0 (\Phi ,  \Omegak  ) .\label{11/24d3} 
\end{equation}
Since $ [ \phi (\mbf{x})^4 , a(f) ]^0    (\Phi ,  \Omegak   )  = -  2 \sqrt{2}(f , \rhobx ) (\Phi ,  \phi (\mbf{x})^3  
 \Omegak  )$, we  have
\begin{align}
[\HI  , a(f) ]^0 (\Phi ,  \Omegak  ) &= \int_{\Rd} \chiIx  [ \phi (\mbf{x})^4 , a(f) ]^0  (\Phi ,   \Omegak  )  \dx 
\notag \\
& = -  2 \sqrt{2} 
\int_{\Rd} \chiIx  (f , \rhobx ) (\Phi ,  \phi (\mbf{x})^3   \Omegak  )   \dx . 
\label{11/24d4} 
\end{align}
By  (\ref{11/24d3}) and (\ref{11/24d4}), we have
\begin{align*}
&\int_{\Rd} f(\mbf{k})^{\ast} ((\Hk - \Ezerok + \omegak ) \Phi , a(\mbf{k})   \Omegak   ) \dk \notag \\
&\qquad \qquad = \int_{\Rd} f(\mbf{k})^{\ast}  \left\{  - 2 \sqrt{2}  \kappa  \frac{\chibk}{\sqrt{\omegak}}\int_{\Rd} \chiIx 
e^{-i \mbf{k}\cdot \mbf{x} } (\Phi ,  \phi (\mbf{x})^3    \Omegak   ) \right\} \dk .
\end{align*}
This yields that 
\[
((\Hk - \Ezerok + \omegak ) \Phi , a(\mbf{k})   \Omegak   ) = 
 -  2 \sqrt{2}  \kappa  \frac{\chibk}{\sqrt{\omegak}}\int_{\Rd} \chiIx 
e^{-i \mbf{k}\cdot \mbf{x} } (\Phi ,  \phi (\mbf{x})^3    \Omegak  )  , \; \; \text{a.e.} \, \Rd . 
\]
Hence, it holds almost everywhere $ \Rd $ that  $a(\mbf{k}) \Omegak   \in \ms{D}(\Hk ) $ and 
\[
(\Hk - \Ezerok + \omegak )  a(\mbf{k})    \Omegak   = 
 - 2 \sqrt{2}  \kappa  \frac{\chibk}{\sqrt{\omegak}}\int_{\Rd} \chiIx   .
\]
Thus the proof is complete. $\square $

\begin{lemma} \label{11/24l2} Suppose \textbf{(H.1)} and \textbf{(H.2)}. 
Let $\Psi \in \ms{D} ( \Hk   ) $. Then, for all $ \epsilon > 0 $  such that $ \epsilon < \frac{1}{c_{\bos} \kappa}$,
\[
 \kappa^2 \int_{\Rd } \chiIx  \chiI (\mbf{x} ' )  \left| (\phi (\mbf{x})^3   \Psi ,  \phi (\mbf{x}')^3 \Psi ) \right| \dx 
 \dx'
 \leq  \lambda_{\epsilon, \kappa} 
\| \Hk   \Psi  \|^2    +  \left( \mu_{\epsilon, \kappa}    
  + \frac{\kappa^2}{2}  \| \chiI \|_{L^1}^2   \right) \| \Psi   \|^2  . 
  \]
\end{lemma}
\textbf{(Proof)}
Let $ \Phi \in \ms{D}_0 $. Since  $[ \phi (\mbf{x}), \, \phi (\mbf{x}' ) ] = 0$, we have 
\begin{align}
 \left| (\phi (\mbf{x})^3   \Phi ,  \phi (\mbf{x}')^3 \Phi ) \right| 
& = \left| (\phi (\mbf{x}')^2 \phi (\mbf{x})^2     \Phi ,  \phi (\mbf{x})   \phi (\mbf{x}')  \Phi ) \right| 
 \notag \\
& \leq \frac{1}{2} \left(   \| \phi (\mbf{x}')^2 \phi (\mbf{x})^2  \Phi   \|^2  
+ \| \phi (\mbf{x}') \phi (\mbf{x})   \Phi   \|^2  \right) \notag \\
& =  \frac{1}{2}   \left(  (\phi (\mbf{x})^4   \Phi ,  \phi (\mbf{x}')^4 \Phi )  
+  (\phi (\mbf{x})^2   \Phi ,  \phi (\mbf{x}')^2 \Phi )  \right)  .  \label{11/24e1} 
\end{align}
Similarly,  we see that 
\begin{equation}
(\phi (\mbf{x})^2   \Phi ,  \phi (\mbf{x}')^2 \Phi )  
\leq \frac{1}{2}   \left(  (\phi (\mbf{x})^4   \Phi ,  \phi (\mbf{x}')^4 \Phi )  
+  (   \Phi ,   \Phi )  \right)  .    \label{11/24e2} 
\end{equation}
By  (\ref{11/24e1}) and (\ref{11/24e2}),  we have
\begin{equation}
\left| (\phi (\mbf{x})^3   \Phi ,  \phi (\mbf{x}')^3 \Phi ) \right| 
\leq (\phi (\mbf{x})^4   \Phi ,  \phi (\mbf{x}')^4 \Phi )  + \frac{1}{2}   (   \Phi ,   \Phi ) .
\label{12/04a}
\end{equation}
By (\ref{12/04a}) and  Corollary \ref{11/24c1}, we have  
\begin{align} 
 \kappa^2 \int_{\Rd } \chiIx  \chiI (\mbf{x} ' )  \left| (\phi (\mbf{x})^3   \Phi ,  \phi (\mbf{x}')^3 \Phi ) \right| \dx   \dx'
&  \leq \| \HIk \Phi \|^2  + \frac{\kappa^2}{2}  \| \chiI \|_{L^1}^2    \, \| \Phi   \|^2  \notag \\
&  \leq   \lambda_{\epsilon, \kappa} 
\| \Hk   \Phi  \|^2    +  \left( \mu_{\epsilon, \kappa}    
  + \frac{\kappa^2}{2}  \| \chiI \|_{L^1}^2   \right) \| \Phi   \|^2  .  \label{12/04b}
\end{align}
Since $\ms{D}_0 $ is a core of $\Hk $,  we see that  
(\ref{12/04b}) holds for all $\Psi \in \ms{D} (\Hk)$. Thus the proof is obtained. $\square $

\begin{proposition} \label{11/24p3}  (Boson number bound) \\
Suppose \textbf{(H.1)} - \textbf{(H.4)}.  For all $ \epsilon > 0 $  such that $ \epsilon < \frac{1}{c_{\bos} \kappa}$, 
\[
( \Omegak  , \Nb \Omegak ) \leq c_{\epsilon,  \kappa }  ,
\]
where $c_{\epsilon , \kappa } = 8  \| \frac{\chib}{\omega^{\, 3/2} }  \|^2   \left(   \lambda_{\epsilon , \kappa} \Ezerok^2
+ \mu_{\epsilon , \kappa}    + \frac{\kappa^2}{2} \| \chiI \|_{L^1}^2  \right)$
\end{proposition}
\textbf{(Proof)} 
Proposition \ref{11/24p2}, we have
\begin{align}
( \Omegak  , \Nb \Omegak    ) 
&= \int_{\Rd } \| a(\mbf{k}) \Omegak    \|^2 \, \dk \notag \\
& \leq 8 \kappa^2   \int_{\Rd } \frac{\chibk^2 }{\omega (\mbf{k})^3 } 
   \left\| \int_{\Rd }  
 \chiIx e^{-i \mbf{k} \cdot \mbf{x}}  \phi (\mbf{x})^3  \Omegak    \dx  \right\|^2   \dk   .
 \label{11/24f1}
\end{align}
By   Lemma \ref{11/24l2} and $\| \Omegak \| =1$,  we have
\begin{align}
\kappa^2 \left\| \int_{\Rd }  
 \chiIx e^{-i \mbf{k} \cdot \mbf{x}}  \phi (\mbf{x})^3 \Omegak    \dx  \right\|^2   
& \leq \kappa^2 \int_{\Rd } \chiIx  \chiI (\mbf{x} ' )  \left| (\phi (\mbf{x})^3  \Omegak ,  \phi (\mbf{x}')^3  \Omegak ) \right| \dx 
 \dx'  \notag \\
& \leq      \lambda_{\epsilon, \kappa} \| \Hk  \Omegak  \|^2    +  \mu_{\epsilon, \kappa}    
  + \frac{\kappa^2}{2}  \| \chiI \|_{L^1}^2       . 
\label{11/24f2} 
\end{align}
By  (\ref{11/24f1})  and  (\ref{11/24f2}), 
\begin{align}
( \Omegak   , \Nb \Omegak   ) & \leq  8    \left\| \frac{\chib }{\omega^{\, 3/2} }  \right\|^2    \left(  
\lambda_{\epsilon, \kappa} \| \Hk  \Omegak  \|^2    +  \mu_{\epsilon, \kappa}    
  + \frac{\kappa^2}{2}  \| \chiI \|_{L^1}^2     \right) \notag \\
 & \leq 8   \left\| \frac{\chib }{\omega^{\, 3/2} }  \right\|^2    \left(  
\lambda_{\epsilon, \kappa} \Ezerok^2    +  \mu_{\epsilon , \kappa}    
  + \frac{\kappa^2}{2}  \| \chiI \|_{L^1}^2     \right) .
\end{align}
Thus   the proof is complete. $\square $


\subsection{Proof of Theorem \ref{MainTheorem}}

\begin{lemma}\label{11/24l1} 
For all $ \Phi \in \ms{D} (\Nb )$ with $\| \Phi \|$=1,
\begin{equation}
(\Phi , P_{0} \Phi )  \; \geq \;  1 -   (\Phi , \Nb \Phi )  . \notag 
\end{equation}
\end{lemma}
\textbf{(Proof)} Since   $ P_0^\bot  = \1 - P_0  $ and    $ ( \Xi , P_0^\bot  \Xi ) \leq  (\Xi , \Nb \Xi ) $ for all $\Xi \in \ms{D}(\Nb )$, the lemma follows. $\square$ 

$\; $ \\
Let  $ \epsilon > 0 $  such that $ \epsilon < \frac{1}{c_{\bos} \kappa}$. Recall that $c_{\epsilon , \kappa } = 8  \| \frac{\chib}{\omega^{\, 3/ 2} }  \|^2   \left(   \lambda_{\epsilon , \kappa} \Ezerok^2
+ \mu_{\epsilon , \kappa}    + \frac{\kappa^2}{2} \| \chiI \|_{L^1}^2  \right)$. 
 We see that 
\begin{equation}
\lim\limits_{\kappa \to 0}  \lambda_{\epsilon , \kappa  }= \lim\limits_{\kappa \to 0}  \frac{1}{1-c_{\bos}\epsilon \kappa} = 1 , \label{11/30a}
\end{equation}
 and 
\begin{equation}
\lim\limits_{\kappa \to 0} \mu_{\epsilon , \kappa  } =\lim\limits_{\kappa \to 0}  \frac{\kappa }{1-c_{\bos} \epsilon \kappa }
  \left (4 d_{\bos} + \frac{c_{\bos}}{4 \epsilon }    \right) = 0 .\label{11/30b}
\end{equation}  
In addition,  Lemma \ref{11/25l1} yields that  
\begin{equation}
 0 \leq \lim\limits_{\kappa \to 0} \Ezerok  \leq   \lim\limits_{\kappa \to 0} \frac{1}{1 + \nu_0 }
\left(  ( \Omegazero , \HI \Omegazero ) \kappa - a \kappa^2 + b \kappa^3 \right)  =  0.
\label{11/30c}
\end{equation}
By (\ref{11/30a}) - (\ref{11/30c}), it follows that
\begin{equation}
\lim_{\kappa \to 0} c_{\epsilon , \kappa } = 0 . \label{11/30d}
\end{equation}

\begin{proposition} \label{11/24p4}  
Suppose \textbf{(H.1)} - \textbf{(H.4)}.  Let  $ \epsilon > 0 $  such that $ \epsilon < \frac{1}{c_{\bos} \kappa}$. Then for sufficiently small $ \kappa >0 $,   it holds that 
\[
 |( \Omega_0 , \Omegak   ) | \geq  \sqrt{ 1- c_{\epsilon , \kappa } } >0. 
\]
In particular
$\Omega_0 $ overlaps with $ \Omegak   $ i.e., $ (\Omega_0 , \Omegak  ) \ne 0 $.
\end{proposition}
\textbf{(Proof)}   Since  dim ker $ (H_0  - E_0   )=1,$ it holds that  $P_0 \Omegak = ( \Omega_0 ,  \Omegak  ) \Omega_0$ and hence 
\[
( \Omegak   , P_0  \Omegak  ) =  |( \Omega_0 , \Omegak   ) |^2. 
\]
By Proposition \ref{11/24p3} and   Lemma \ref{11/24l1},
\[
( \Omegak  , P_0  \Omegak   ) \geq 1 - ( \Omegak   , \Nb  \Omegak   )\geq 
1- c_{\epsilon , \kappa } . 
\]
By (\ref{11/30d}), it holds that $1- c_{\epsilon , \kappa }>0 $ for sufficiently small $ \kappa >0 $.  Thus the proof is complete. $\square $

$\; $ \\
\textbf{(Proof of Theorem \ref{MainTheorem})} \\
We show that the conditions \textbf{(A.1)} -  \textbf{(A.4)} in Appendix are satisfied. 
By \textbf{(H.1)} - \textbf{(H.3)}, we see that  \textbf{(A.1)} and \textbf{(A.2)} are satisfied.
Note that  $\sigma_{\textrm{p}} \left(   H_0^{\bot} \right) = \{ \emptyset \}  $. Then,   Proposition \ref{11/24p4} yields that \textbf{(A.3)} follows.
Let 
\[
\tilde{ \Omega }_\kappa  = \frac{1}{(\Omegazero ,   \Omegak  ) }  \Omegak .
\] 
By Remark A in  Appendix,  it is enough to show that 
\[
\lim_{\kappa \to 0} \|  \tilde{ \Omega }_{\kappa} \| =1 ,
\] 
and then, \textbf{(A.4)} holds.
Since $\Omegak $ is normalized, we  show that 
$ \lim\limits_{\kappa \to 0} | (\Omegazero , \Omegak ) |  = 1 $.  
By Proposition \ref{11/24p4} and (\ref{11/30d}),  we have
\[
1 \geq \lim_{\kappa \to 0}  |(\Omega_0 , \Omegak ) |  \geq \lim\limits_{\kappa \to 0}  \sqrt{ 1- c_{\epsilon , \kappa } }=1.
\]
Thus,  the proof is complete. $\square $ \\


$\; $ \\  
{\Large {Appendix}(\cite{Ar14})} \\

Let  $S_0$ and $\SI $ be linear operators on  a complex Hilbert space $\ms{H}$. Let
\[
\qquad  \Sk  = S_0 + \kappa \SI  ,  \qquad \kappa  \in\mbb{R} .
\]
Suppose the following conditions. \\

\textbf{(A.1)} The operators $S_{0}$ and  $ \SI $ are symmetric. 

\textbf{(A.2)} The operator $S_{0}$ has a simple eigenvalue  $E  $. 

$\;$ \\
Let $ \Psi_E $ be a normalized eigenvector of $S_0$ with respect to $ E  $ ;
\[
\qquad  S_{0} \Psi_E = E \Psi_E , \qquad \| \Psi_E  \| = 1 .
\]
Let $P_{E}$ be the projection onto the closed subspace $ \ms{M}_{E}  = \left\{z \,  \Psi_E \left| \frac{}{} \right. z \in \mbb{C} \right\}$.  Let $ P_{E}^{\bot} = \1 - P_{E}$. 
Since $S_{0} $ is symmetric, $S_{0}$ is reduced by $  \ms{M}_E  $ and $  \ms{M}^{\bot}_E  $, respectively. Let  
$ S_{0,E}^{\bot} = S_{0 \, \restr \ms{D}(S_{0})  \cap \ms{M}_E^{\bot}}$. \\
$\;$ \\
For a symmetric operator $S$, we set 
\[
\ms{E}_0 (S ) = \inf_{\Psi \in \ms{D}(S) , \| \Psi \|=1} (\Psi , S \, \Psi ) . 
\]
$\;$ \\
\textbf{Proposition A (\cite{Arai}, Theorem 2.7)}
Suppose \textbf{(A.1)} and \textbf{(A.2)}.  Assume that $S_{0}$ is self-adjoint,  $E=E_{0}$ where $ E_{0} = \inf \sigma (S_0 ) $, and
\[
\Psi_{E_0}  \in \ms{D} (\SI (S_{0 , E_0 }^{\bot} - E_0  )^{-1}   P_{E_0 }^{\bot}\SI ).
\] 
Then for all $\kappa \in \mbb{R}$, 
\[
\ms{E}_0 (\Sk ) \leq  E_0 + \frac{1}{1 + \nu_0 }
\left(  ( \Psi_{E_0} , \SI \Psi_{E_0} ) \kappa - a \kappa^2 + b \kappa^3 \right) ,
\]
where 
\begin{align*}
& \nu_0 = \|  ( S_{0,E_0}^{\bot} -  E_0  )^{-1} P_{E_0 }^{\bot}  \SI \Psi_{E_0}  \|^2 , \\ 
&  a = ( P_{E_0 }^{\bot} \SI \Psi_{E_0}  , \,  (   S_{0,E_0}^{\bot}- E_0 )^{-1} P_{E_0 }^{\bot}
 \SI \Psi_{E_0}   )  , \\
& b=( (S_{0,E_{0}}^{\bot}  - E_0 )^{-1}   P_{E_0 }^{\bot} \SI  \Psi_{E_0} , \, \SI (
 S_{0,E_0}^{\bot} - E_0  )^{-1} P_{E_0 }^{\bot}  \SI \Psi_{E_0})  .
\end{align*}
$\; $ \\
Let  $\Psi $ and $\Phi $  be vectors in a Hilbert space. We say that $\Psi $ overlaps with  $\Phi $ 
if $ (\Psi , \Phi) \ne 0$.   \\

\textbf{(A.3)} There exists constants $r>0 $ such that for all $\kappa \in (-r , 0) \cup ( 0,  r) $, 
  $\Sk $ has an eigenvalue $\Ek $ such that $ \Ek  \notin \sigma_{\textrm{p}} \left(   S_0^{\bot} \right) $ and $\Psi_E $ overlaps  with a vector in  ker $ ( \Sk  - \Ek  )$. 

$\;$\\
Under the conditions \textbf{(A.1)} - \textbf{(A3)}, it follows (\cite{Arai}, Proposition 2.1)    that for each $ \kappa \in (-r ) \cup ( 0,  r) $, there exists a non-zero
vector $\PsiEk \in $ ker $( \Sk -\Ek )$ such that
\begin{align}
& \Ek = E + \kappa (\Psi_E , \SI \PsiEk ) , \tag{A.1} \\
& \PsiEk= \Psi_E + \Phik , \tag{A.2}
\end{align}
where $\Phik = - \kappa   (S_0^{\bot} - \Ek )^{-1} P_0^{\bot} \SI \PsiEk  $.

$\; $ \\
\textbf{Remark A} Consider the case of  dim ker $(\Sk - \Ek ) =1$. Let $ \Xi_{\Ek}    $ be the normalized vector in ker $(\Sk - \Ek  )$. Then it holds that 
$\PsiEk  = \frac{1}{(\Psi_{E}   ,   \Xi_{\Ek}) }\Xi_{\Ek}  . $ \\

\textbf{(A.4)}  $\lim\limits_{\kappa \to 0}    \|  \PsiEk \| = 1 $. \\

$\;$ \\
\textbf{Theorem A (\cite{Ar14}; Theorem 3.1)} \\
 Suppose \textbf{(A.1)} -  \textbf{(A.4)}. Then  it holds that 
 \[
 \Ek    = E + \kappa (\Psi_E , S_{\I} \Psi_E ) + o (\kappa ) . 
 \]

$\; $ \\
{\large {\textbf{Acknowledgments}}} \\
It is a pleasure to thank Professor  Fumio Hiroshima for his comment and  advice.
This work is supported by JSPS  Grant $20$K$03625$.

\end{document}